\documentclass[12pt]{article}
\usepackage{amsfonts}
\usepackage{amsmath}
\usepackage{amssymb}
\usepackage{mathptmx}
\usepackage[dvips]{color}
\usepackage{epsfig}
\usepackage{graphicx}
\begin{document}
\def\be{\begin{eqnarray}}
\def\en{\end{eqnarray}}
\def\non{\nonumber}
\def\A{{ A}}
\def\ov{\overline}
\def\up{\uparrow}
\def\dw{\downarrow}
\def\vp{\varepsilon}
\title{Understanding $B\to J/\psi \phi$ in the Standard Model}

\author{Ying Li\footnote{e-mail: liying@ytu.edu.cn.}
 \\
{\small \it  Department of Physics, Yantai University, Yantai
264005, China}
\\~~\\Hai-Yang Cheng
 \\
{\small\it Institute of Physics, Academia Sinica, Taipei, Taiwan
115, R.O.C.}}
 \maketitle
\begin{abstract}
The rare decay $\bar B_d^0\to J/\psi \phi$ can proceed via four
distinct mechanisms: (i) production of the $\phi$ via tri-gluon
fusion, (ii) photoproduction of the $J/\psi$ or $\phi$, (iii)
final-state rescattering of $D_s^{(*)}D_s^{(*)}$ produced in the
$\bar B_d$ decay to $J/\psi\phi$, and (iv) production of the $\phi$
via $\omega-\phi$ mixing. In this work, we examined the
contributions of photoproduction and final-state rescattering to
$\bar B_d^0\to J/\psi \phi$ and found that the corresponding
branching ratios were of the orders $10^{-11}$ and $10^{-9}$,
respectively. Hence, this decay is dominated by the $\omega-\phi$
mixing effect.

\end{abstract}
\par {\bf 1.}
The observation of $B$ decays to charmonium provides important
evidence for the Cabbio-Kabayashi-Maskawa  model, as well as an
important advance in our understanding of the Standard Model and QCD
dynamics. Recently, Belle  reported an upper limit $9.4 \times
10^{-7}$ for the branching ratio of $B^0\to J/\psi \phi$ at the 90\%
confidence level \cite{Liu:2008bt}. This process is expected to be
suppressed by the Okubo-Zweig-Iizuka (OZI) rule \cite{Okubo:1963fa}
disfavoring disconnected quark diagrams.

\begin{figure}[tp]
\begin{center}
\includegraphics[width=0.6\textwidth]{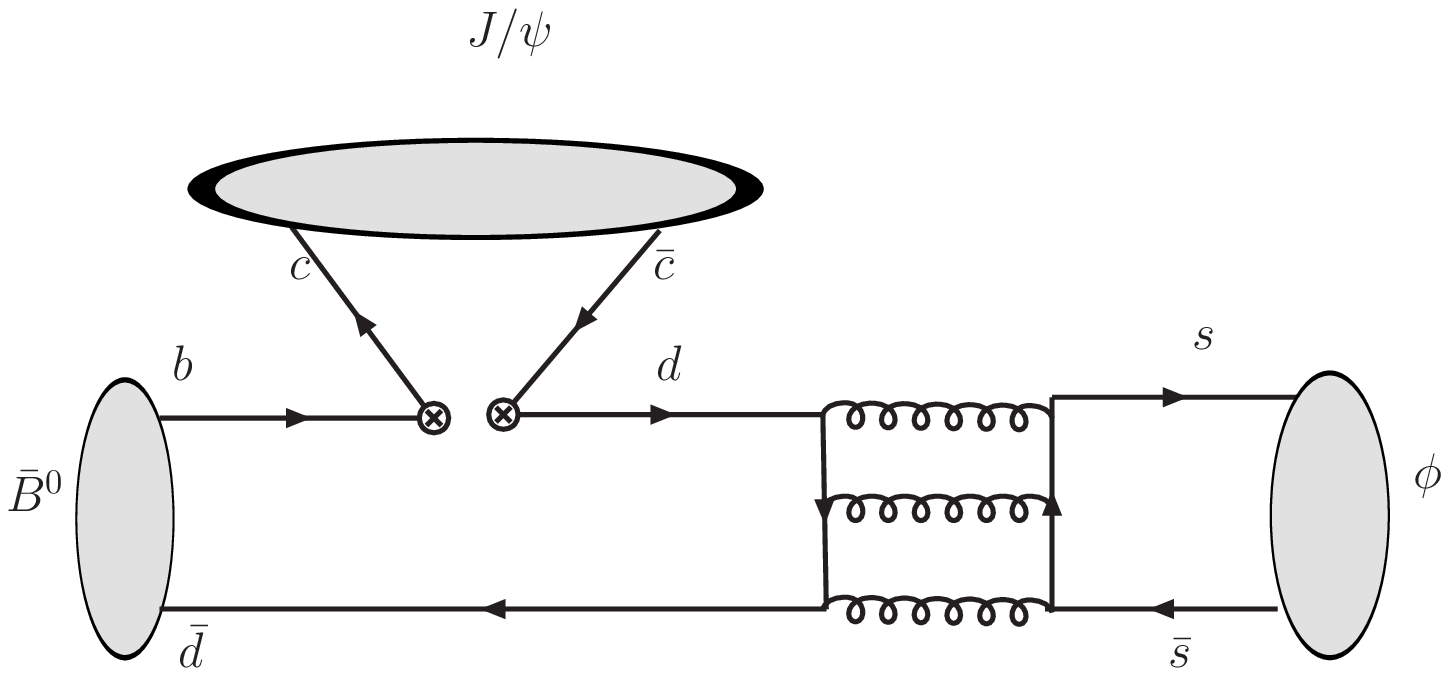}
\caption{ Quark-level diagram for the $B^0 \to J/\psi \phi$ decay
via tri-gluon exchange.} \label{fig:trigluon} \end{center}
\end{figure}

\begin{figure}[t]
\begin{center}
\includegraphics[width=0.6\textwidth]{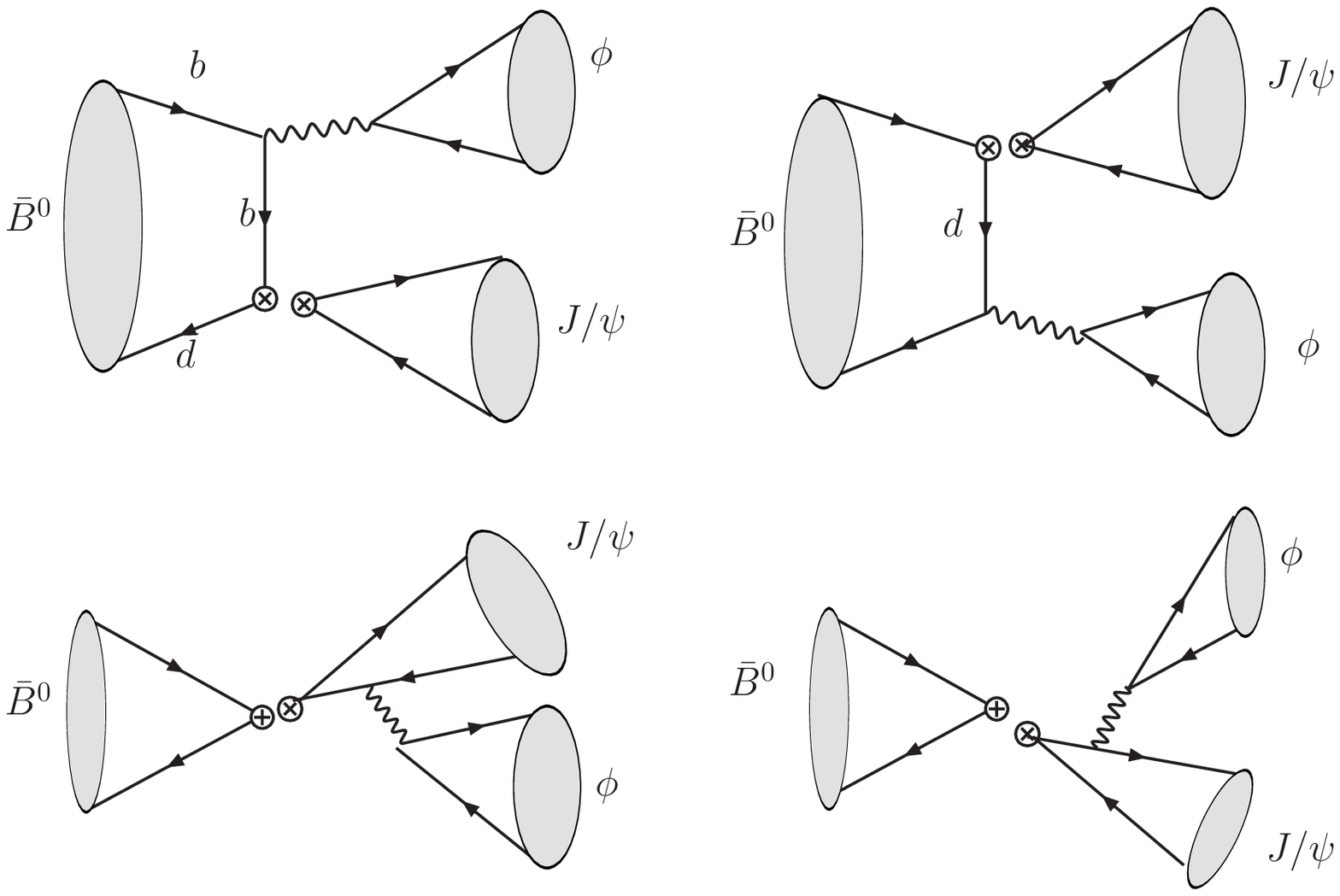}
\caption{ Quark-level diagrams for the $B^0 \to J/\psi \phi$ decay
produced in the photo-production mechanism. }
\label{fig:photoproduction} \end{center}
\end{figure}

\begin{figure}[t]
\begin{center}
\includegraphics[width=0.6\textwidth]{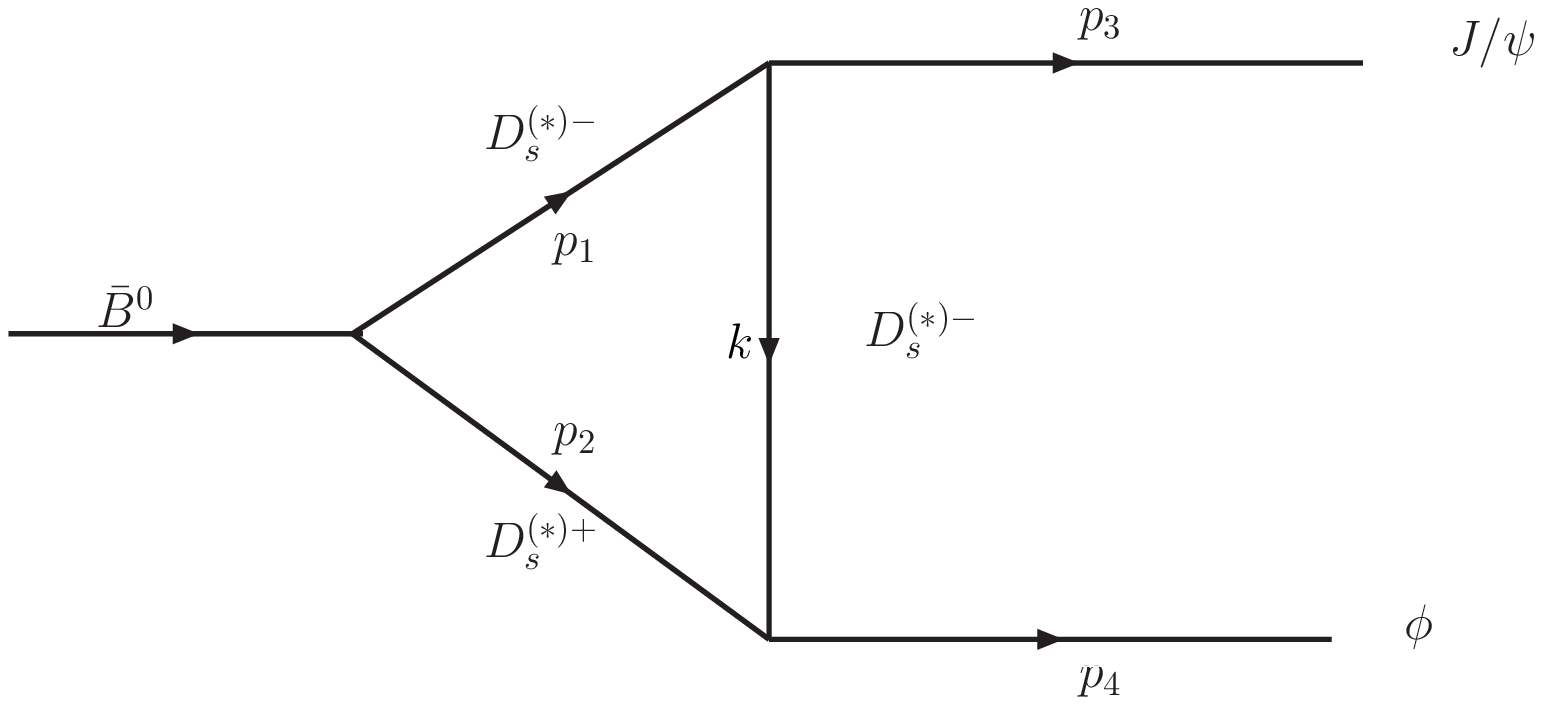}
\caption{Long-distance contribution to $\overline B^0\to J/\psi\phi$
} \label{fig:fsi}
\end{center}
\end{figure}

The main processes for $\bar B_d^0\to J/\psi \phi$ can be sorted
into four different classes: (i) the neutral vector meson $\phi$ is
produced through tri-gluon fusion (Fig.~\ref{fig:trigluon}), which
is formally the reason why this channel is OZI-suppressed, (ii) the
$J/\psi$ or $\phi$ arises from a photon emission, followed by
fragmentation (Fig. \ref{fig:photoproduction}), (iii) the decay
particles $J/\psi$ and $\phi$ are produced through long-distance
final-state interactions (FSI) (see Fig. \ref{fig:fsi}), and (iv)
the $\phi$ comes from the decay of $B\to J/\psi\omega$ followed by
$\omega-\phi$ mixing; that is, $\phi$ is not a pure $s\bar s$ state
and contains a tiny $q\bar q$ component.

In \cite{Gronau:2008kk}, Gronau and Rosner pointed out that the
major contribution to the decay $\bar B_d^0\to J/\psi \phi$ arises
from $\omega-\phi$ mixing. Neglecting isospin violation and the
admixture with the $\rho^0$ meson, one can parameterize
$\omega$--$\phi$ mixing in terms of an angle $\delta$ such that the
physical $\omega$ and $\phi$ are related to the ideally mixed states
$\omega^I \equiv (u \bar u + d \bar d)/\sqrt{2}$ and $\phi^I \equiv
s \bar s$ by
\begin{eqnarray}\label{mixing}
\left(
\begin{array}{c} \omega \\ \phi \end{array} \right) = \left(
\begin{array}{c c} \cos \delta & \sin \delta \\ - \sin \delta & \cos
\delta
\end{array} \right)
\left( \begin{array}{c} \omega^I \\ \phi^I \end{array} \right),
\end{eqnarray}
and the mixing angle is approximately $\delta = - (3.34 \pm
0.17)^\circ$ \cite{Benayoun:1999fv}. Within this mechanism, the
authors estimated the rates of this decay mode and other similar
processes in $B^0$ and $B_s^0$ decays, and found that the Belle's
upper limit is about a factor of five above their estimation. Also,
they argued that the final-state rescattering contributions to this
decay mode are very small and can be neglected.

Let us make crude estimates of the various contributions to $B\to
J/\psi \phi$ by the aforementioned four mechanisms. Due to the
complicated QCD dynamics, it is difficult to calculate the
 tri-gluon fusion reliably. Roughly, the tri-gluon fusion contribution gives
\begin{eqnarray}
{\cal B}(\bar B^0\to J/\psi\,\phi)_{\rm tri-gluon}={\cal B}(\bar
B^0\to J/\psi\,\omega) \alpha_s^3\approx 2.7\times 10^{-5}\times
(0.3)^3 \sim 2.4\times 10^{-8},
\end{eqnarray}
where use of ${\cal B}(\bar B^0\to J/\psi\,\omega)\approx { \cal
B}(\bar B^0\to J/\psi\rho)=(2.7\pm0.4)\times 10^{-5}$ \cite{PDG} has been made.
The contribution of photoproduction is calculable to the leading
power of the $1/m_b$ expansion and is of order
\begin{eqnarray}
{\cal B}(\bar B^0\to J/\psi\phi)_{\rm photoproduction}={\cal B}(\bar
B^0\to J/\psi\gamma) \alpha_{\rm em}^2\sim 10^{-7}\times (1/137)^2
\sim 10^{-11}.
\end{eqnarray}
In the final-state rescattering picture, the $B\to J/\psi \phi$
decay proceeds via a $B$ meson  decay into $D_s^{(*)+}D_s^{(*)-}$
through $W$-exchange followed by a rescattering of
 $D_s^{(*)+}D_s^{(*)-}$ to $J/\psi \phi$ through
$D_s^{(*)\pm}$ exchange. It is anticipated that
\begin{eqnarray}
{\cal B}(\bar B^0\to J/\psi\phi)_{\rm FSI}={\cal B}(\bar B^0\to
D_s^{(*)+}D_s^{(*)-}) (10^{-3}-10^{-4})\sim 3\times 10^{-8}-3\times
10^{-9},
\end{eqnarray}
where the analysis of final-state interactions in $B\to \phi K^*,
\rho K^*$ suggests that the rate of the $B$-meson decay into the
final state under consideration (for example, $\bar B^0\to J/\psi \phi$) is
suppressed relative to that of the intermediate state ($\bar B^0\to
D_s^{(*)+}D_s^{(*)-}$ in this example) by three to four orders of
magnitude \cite{CCSfsi}. Finally, the production of $J/\psi\phi$ through
$\omega-\phi$ mixing is expected to be
\begin{eqnarray}
{\cal B}(\bar B^0\to J/\psi\phi)_{\rm \omega-\phi~ mixing}={\cal B}
(\bar B^0\to J/\psi\,\omega)\sin^2\delta \approx 2.7\times
10^{-5}\times (0.08)^2 \sim 1.7\times 10^{-7}.
\end{eqnarray}
Therefore, the rare decay $B\to J/\psi\phi$ is indeed dominated by
the $\omega-\phi$ mixing effect.

In this letter, we will study the effects of photoproduction and
final-state rescattering in more detail even though they are not the
main contributions to $B\to J/\psi\,\phi$. We wish to have
quantitative results to confirm the above crude estimates.

\vskip 0.4cm {\bf 2.}~ Firtst, Let us evaluate the photoproduction,
which plays an important role in decay modes such as $B\to \rho
K^*,\, \rho\phi$ \cite{Beneke:2005we}. In this mechanism, $\bar
B_d^0\to J/\psi\phi$ can be regarded as the cascade process $\bar
B_d^0\to J/\psi\gamma\to J/\psi \phi$ or $\bar B_d^0\to
\phi\gamma\to J/\psi \phi$. The radiative decay $B\to V\gamma$ has
been well studied in the frameworks of the QCD factorization
approach \cite{Bosch:2001gv}, the perturbative QCD approach (pQCD)
\cite{Wang:2007an} and soft-collinear effective theory
\cite{SCETVgamma}. Due to the suppression of Wilson coefficients, we
will neglect the contribution from $\bar B_d^0\to \phi\gamma\to
J/\psi \phi$.

According to the Feymann diagrams depicted in Fig.~\ref{fig:photoproduction},
the amplitude of $\bar B_d^0\to J/\psi\phi$ can be written as
\begin{eqnarray}
   {\cal  M}&\simeq& {\cal A}^\mu(\bar B_d^0\to
   J/\psi\gamma)\frac{-ig_{\mu\nu}}{q^2}\left(-\frac{1}{3}e\right)\langle0|\bar
   s\gamma^\nu s|\phi\rangle \nonumber\\
   &\simeq&\left(\frac{f_\phi
  \sqrt{4\pi\alpha_{em}}}{3m_\phi}\right){\cal A}^\mu(\bar B_d^0\to
   J/\psi\gamma)\varepsilon_{\mu}^{*},
\end{eqnarray}
where we have used $\langle 0|\bar{s}\gamma^\nu s|\phi\rangle=
-m_\phi f_\phi\varepsilon^{\nu*}$ and $f_\phi$ and $m_\phi$ are the
decay constant and mass of the $\phi$ meson, respectively.
Therefore, we obtain the result
\begin{eqnarray}
   &&{\cal B}(\bar B_d^0\to J/\psi \phi)\simeq R_\phi {\cal B}(\bar B_d^0\to J/\psi
   \gamma),\nonumber\\
   && R_\phi=\left\vert\frac{f_\phi\sqrt{4\pi\alpha_{em}}}{3m_\phi} \frac{
  }{}\right\vert^2\simeq 0.0003\,,
\end{eqnarray}
with $f_\phi=0.237\,\mathrm{GeV}$. In the literature, it has been
estimated that ${\cal B}(\bar{B}^0_d \to J/\psi\gamma)
=7.7\times10^{-9}$ \cite{Lu} in QCD factorization and ${\cal
B}(\bar{B}^0_d \to J/\psi\gamma)=4.5\times10^{-7}$ \cite{Li:2006xe}
in perturbative QCD. Therefore, the predictions of QCDF and pQCD
differ by one to two orders of magnitude.The possible reason for
this huge discrepancy was explained in Ref.\cite{Li:2006xe}. Roughly
speaking, this is due mainly to the use of different $J/\psi$ wave
functions in Ref.\cite{Lu} and Ref.\cite{Li:2006xe}. If the charm
quark is heavy, the wave function of $J/\psi$ will be symmetric
under $x\leftrightarrow 1-x$ and sharply peaked around $x= 0.5$.
However, the cross section of $e^+e^-\to\eta_c+J/\psi$ calculated
within the NRQCD approach is much smaller than the experimental
data. Bondar and Chernyak \cite{Bondar} have pointed out that the
origin of the discrepancy is due to the fact that the charm quark is
not heavy enough and, as a result, the charmonium wave functions are
not sufficiently narrow for a reasonable application of NRQCD to the
description of charmonium production. Using more realistic models,
these authors have proposed a new wave function for  $J/\psi$, which
can be used to explain the data well. This new wave function is
employed in Ref. [12], while the delta function is used in Ref.
[11].

Even taking the pQCD result for $\bar{B}^0_d \to J/\psi\gamma$, the
photoproduction mechanism leads to a very small branching ratio for
$\bar B_d^0\to J/\psi\phi$ of order $10^{-11}$, which is not
accessible even at the future Super-B factories. Since the $\phi$ is
produced from a virtual photon which is transversely polarized
mostly, the longitudinal polarization of the decay $B \to J/\psi
\phi$ via photoproduction will be very small.

\vskip 0.4cm {\bf 3.}~ As mentioned above, $\bar B^0\to J/\psi\phi$
receives long-distance contributions from a $B$ meson decay into
$D_s^{(*)+}D_s^{(*)-}$ followed by a rescattering of
$D_s^{(*)+}D_s^{(*)-}$ to $J/\psi \phi$. The $D_s^{(*)} D_s^{(*)}$
states from $\ov B^0$ decays can rescatter to $J/\psi\,\phi$ through
the $t$-channel $D_s^{(*)}$ exchange in the triangle diagrams
depicted in Fig.~\ref{fig:fsi}. Before proceeding, we would like to
remark briefly on the motivation for considering the rescattering
mechanism with $D_s^{(*)}$ exchange. At the hadron level,
final-state interactions manifest as the rescattering processes with
$s$-channel resonances and one particle exchange in the $t$-channel.
Due to the lack of the existence of resonances at energies close to
the $B$ meson mass, we will therefore model FSIs as rescattering
processes of some intermediate two-body state with one particle
exchange in the $t$-channel. we will compute the absorptive part via
the optical theorem \cite{CCSfsi}. We consider charm intermediate
states based on the idea that if the intermediate states are CKM
more favored than the final state, then the absorptive part of the
final-state rescattering amplitude can easily give rise to large
strong phases and make significant contributions to the rates. It
has been shown in Ref.\cite{CCSfsi} that the direct $CP$-violating
partial rate asymmetries in charmless $B$ decays to $\pi\pi/\pi K$
and $\rho\pi$ are significantly affected by final-state rescattering
and their signs are generally different from those predicted by the
short-distance approach. Especially, the calculated {\it CP}
asymmetry $A_{CP}(K^+\pi^-)=-0.14^{+0.01}_{-0.03}$ for $B^0\to
K^+\pi^-$ via rescattering \cite{CCSfsi} agrees with experiments in
both magnitude and sign, whereas the QCD factorization prediction
$A_{CP}(K^+\pi^-)\approx 0.045$ \cite{BN} is wrong in sign. This
example illustrates that the rescattering approach gives a
reasonable description of FSIs.

To evaluate Fig.~\ref{fig:fsi}, we note that the effective
Lagrangian for $\phi D_s^{(*)}D_s^{(*)}$ vertices can be found in
\cite{CCSfsi}, and the effective Lagrangian for $J/\psi
D_s^{(*)}D_s^{(*)}$ vertices is given by
\begin{eqnarray}
\mathcal{L}_{\psi D_sD_s}&=&i g_{\psi D_sD_s} \psi_\mu
\left(\partial^\mu
D_s{D_s}^{\dagger}-D_s\partial^\mu {D_s}^{\dagger}\right),\label{LpsiDD}\\
\mathcal{L}_{\psi D_s^*D_s}&=&\!\!-2f_{\psi\!
D_s^*\!D_s}\varepsilon^{\mu\nu\alpha\beta}\partial_\mu \psi_\nu
\!\!\left(\partial_\alpha {D_s}^*_\beta {D_s}^{\dagger}\!\!
+ \!\! D_s \partial_\alpha {D_s}^{*\dagger}_\beta\!\! \right)\!,\label{LpsiD*D}\\
\mathcal{L}_{\psi D_s^*D_s^*}&=&-i g_{\psi D_s^* D_s^*} \Bigl\{
\psi^\mu
\left(\partial_\mu D_s^{*\nu} {D_s}_\nu^{*\dagger} -D_s^{*\nu}\partial_\mu {D_s}_\nu^{*\dagger} \right)\nonumber\\
&&+ \psi^\nu D_s^{*\mu}\partial_\mu{D_s}^{*\dagger}_{\nu} -
\psi_\nu\partial_\mu D_s^{*\nu}  {D_s}^{*\mu\dagger}  \mbox{}
\Bigr\}.\label{LpsiD*D*}
\end{eqnarray}
The coupling constants for the $\phi D_s^{(*)}D_s^{(*)}$ vertices can
be related to the parameters $g_V$, $\beta$ and $\lambda$ appearing
in the effective chiral Lagrangian describing the interactions of
heavy mesons with low momentum vector mesons
\cite{Casalbuoni} in the following manner
\begin{eqnarray}\label{parameter}
g_{\phi D_sD_s}=\frac{\beta\,g_V}{\sqrt{2}}=3.75\,, &&
f_{\phi D_sD_s^*}=\frac{\lambda\,g_V}{\sqrt{2}}=2.30\,{\rm GeV}^{-1}, \nonumber\\
f_{\phi D_s^*D_s^*}=\frac{\lambda\,g_V}{\sqrt{2}}m_{D_s^*}=4.85 \,,
&& g_{\psi D_sD_s}=4f_{\psi D_s^*D_s}=4f_{\psi D_s^*D_s^*}/m_{D_s^*}=10\,,
\end{eqnarray}
where we have assumed $\beta=0.9$ and $\lambda=0.56$~GeV$^{-1}$
\cite{Isola2003} and the relation $g_V=m_\rho/f_\pi$
\cite{Casalbuoni}. The couplings for $J/\psi D_s^{(*)}D_s^{(*)}$ are
taken from Ref.\cite{Deandrea:2003pv} based on an effective field
theory of quarks and mesons. Note that the same $\phi
D_s^{(*)}D_s^{(*)}$ vertex also appears in the rescattering
contribution to $B\to \phi K^*$. A study in \cite{CCSfsi} shows that
the rescattering mechanism via $D_s^{(*)}$ exchange can enhance the
rate and yield a large transverse polarization in $B\to \phi K^*$.

In total, there are eight different FSI diagrams in
Fig.~\ref{fig:fsi}. The $\overline B^0\to D_s^{(*)-} D_s^{(*)+}\to
J/\psi\phi$ amplitudes via $D_s^{(*)}$ exchange are similar to the
$\overline B\to \bar D_s^{(*)} D^{(*)}\to \bar K^*\phi$ amplitudes
via  $D_s^{(*)}$ exchange that have been studied in
Ref.\cite{CCSfsi}. Therefore, the amplitudes of the former can be
obtained from the latter through the replacements $\bar K^*\to
J/\psi$ and $D^{(*)}\to D_s^{(*)}$. For example, the absorptive part
contributions of $\overline B^0\to D_s^{-} D_s^{+}\to J/\psi\phi$
amplitudes via $D_s$ exchange is given by
 \be \label{eq:DsDDspre}
 \A bs\,(D_s^-D_s^+;D_s) &=& {1\over 2}\int {d^3\vec p_1\over (2\pi)^32E_1}\,{d^3 \vec p_2\over
(2\pi)^3 2E_2}\,(2\pi)^4\delta^4(p_B-p_1-p_2)A(\ov B^0\to D_s^- D_s^+) \non \\
 &&\times  (2i) g_{D_s D_s\phi} {F(p_1,k)F(p_2,k)\over
 t-m_{D_s}^2}(-2i)g_{D_s D_s J/\psi}\,
 (\vp^*_3\cdot p_1) (\vp^*_4\cdot p_2),
 \en
where $k=p_1-p_3=p_4-p_2$ is the momentum of the exchanged particle.
Since the particle exchanged in the $t$ channel is off shell and
since final state particles are hard, form factors or cutoffs must
be introduced to the strong vertices to render the calculation
meaningful in perturbation theory. The form factor $F(p,k)$ for the
off-shell effect of the exchanged particle can be parametrized as
 \be \label{FF}
 F(p,k)=F(t,m_{\rm exc})=\,\left({\Lambda^2-m_{\rm exc}^2\over \Lambda^2-t}\right)^n,
 \end{eqnarray}
normalized to unity at $t=m_{\rm exc}^2$, where $m_{\rm exc}$ is the
mass of the exchanged particle. The cutoff $\Lambda$ in the form
factor $F(t)$  should be not far from the physical mass of the
exchanged particle. To be specific, we write \cite{CCSfsi}
\begin{equation} \label{Lambda}
\Lambda=m_{\rm
exc}+\eta\Lambda_{\rm QCD},
\end{equation}
where the parameter $\eta$ is expected
to be of order unity and it depends not only on the exchanged
particle but also on the external particles involved in the
strong-interaction vertex. As we do not have first-principles
calculations for form factors, we shall use the measured decay rates
to fix the unknown cutoff parameters. Although the strong couplings
are large in magnitude, the rescattering amplitude is suppressed
by a factor of $F^2(t)\sim (m^2\Lambda_{\rm QCD}^2/t^2)^n$.
Consequently, the off-shell effect will render the perturbative
calculation meaningful. It is also evident from Eq.
(\ref{eq:DsDDspre}) that the final-state rescattering contributions
vanish in the heavy quark limit, as it should be.

As discussed in Ref.\cite{CCSfsi}, the FSI contribution from the
$\ov B\to D_s^- D_s^+$ decay will affect both $A_L$ and
$A_\parallel$ amplitudes of the $\ov B\to J/\psi\phi$ decay, whereas
both $\ov B\to D^*_s D_s$ and $\ov B\to D_s D_s^*$ will affect only
the $A_{\bot}$ term of the $\ov B\to J/\psi\phi$ decay amplitude.
Finally, the FSI effect from the decay $\overline B\to D^*_s D_s^*$
contributes to all three polarization components
$A_{L,\parallel,\bot}$.

In order to perform a numerical study of the long-distance
contributions, we need to specify the short-distance $A(\ov B^0\to
D_s^{(*)-} D_s^{(*)+})$ amplitudes. This decay proceeds only through
$W$-exchange, and it can be calculated in pQCD effectively without
introducing any new parameters \cite{Li2005,Li2007}. Numerically, we
have (in units of $V_{cb}V_{cd}^*$~GeV)
\begin{eqnarray}
{\cal A}(B^0\to D_s^{+}D_s^{-}) &=& 7.93\times10^{-6} + i0.94\times 10^{-6}, \nonumber \\
{\cal A}(B^0\to D_s^{*+}D_s^{-}) &=& 0.98\times 10^{-6} + i
1.12\times 10^{-7},
\end{eqnarray}
and
\begin{eqnarray}
a &=& 1.9\times 10^{-6} - i1.4\times 10^{-7}, \nonumber \\
b &=& -6.5\times 10^{-9} + i4.7\times 10^{-8}, \nonumber \\
c &=& 6.7\times 10^{-9} - i4.9\times 10^{-8},
\end{eqnarray}
for the $B^0\to D_s^{*+}D_s^{*-}$ amplitude given by
\begin{eqnarray}
{\cal A}(B\to
D_s^{*+}(p_1,\varepsilon_1)D_s^{*-}(p_2,\varepsilon_2))=a
(\varepsilon_1^*\cdot\varepsilon_2^*)+b (\varepsilon_1^*\cdot
p_2)(\varepsilon_2^*\cdot
p_1)+ic\epsilon_{\alpha\beta\mu\nu}\varepsilon_1^{*\alpha}\varepsilon_2^{*\beta}
p_1^\mu p_2^\nu.
\end{eqnarray}
It follows that the branching ratios of $B\to D_s^{(*)+}D_s^{(*)-}$ read
\begin{eqnarray}\label{DDresults}
&{\cal B}(B^0\to D_s^{+}D_s^{-})=(3.3\pm 1.1)\times 10^{-5}, \nonumber\\
&{\cal B}(B^0\to D_s^{*+}D_s^{-})=(2.6\pm 1.0)\times 10^{-5}, \nonumber\\
&{\cal B}(B^0\to D_s^{*+}D_s^{*-})=(1.2\pm 0.4)\times 10^{-5}.
\end{eqnarray}
In the above calculation, we have included the errors coming from
the the hadronic wave functions that are dominated by the
$D_s^{(*)}$ meson distribution amplitude rather than the $B$ meson,
as the latter is more or less fixed by the well measured channels
such as $B\to K\pi, \pi\pi$. Since we employ the updated $D_s^{(*)}$
distribution amplitude \cite{dwavefunc}
\begin{eqnarray} \label{dwavefunc}
\phi_{D_s^{(*)}}(x,b)=\frac{3}{\sqrt{6}}f_{D_s^{(*)}}x(1-x)\left[1+a_{D_s^{(*)}}(1-2x)\right]\exp
\left (\frac{-\omega^2 b^2 }{2}\right ),
\end{eqnarray}
with $a_{D_s^{(*)}}=0.5\,\mathrm{GeV}$ and
$\omega=(0.6-0.8)\,\mathrm{GeV}$, our predictions are slightly
smaller than the ones in \cite{Li2005} \footnote{Our estimate of
${\cal B}(B^0\to D_s^+D_s^-)$ is smaller by more than a factor of
two than a value of $(7.8^{+2.0}_{-1.6})\times 10^{-5}$ obtained in
\cite{Li2005} using the same PQCD approach. This is mainly due to
the additional exponential term exp$(-\omega^2 b^2/2)$  in the
revised $D_s^{(*)}$ distribution amplitude, Eq. (\ref{dwavefunc}).
Based on the diagrammatic approach, an estimate of ${\cal B}(B^0\to
D_s^+D_s^-)=(4.0^{+1.8}_{-1.5})\times 10^{-6}$  was obtained in Ref.
\cite{Gronau:BtoDD}, which is smaller than the PQCD result by one
order of magnitude. This should be checked by experiment.} but
consistent with the current experimental limits \cite{BaBar, Belle}
\begin{eqnarray}
&{\cal B}(B^0\to D_s^{+}D_s^{-})& <1.0\times 10^{-4}~({\rm BaBar}), ~~<3.6\times 10^{-5}~({\rm Belle}),  \nonumber\\
&{\cal B}(B^0\to D_s^{*+}D_s^{-})& <1.3\times 10^{-4}~({\rm BaBar}),  \nonumber\\
&{\cal B}(B^0\to D_s^{*+}D_s^{*-})& <2.4\times 10^{-4}~({\rm
BaBar}).
\end{eqnarray}

For the parameter $\eta$ in Eq. (\ref{Lambda}), we shall use the one
$\eta=0.80$ extracted from $B\to \phi K^*$ \cite{CCSfsi}. With the
$B\to D_s^{(*)+}D_s^{(*)-}$ amplitudes given before and the
parameters (\ref{parameter}), the decay rate and the longitudinal
polarization fraction $f_L$ of $B\to J/\psi\phi$ due to final-state
rescattering turn out to be
\begin{eqnarray} \label{BRtheory}
{\cal B}(\bar{B}^0 \to J/\psi\phi)_{\rm FSI}
&=&(3.7^{+5.8}_{-2.5})\times 10^{-9}, \qquad f_L = 0.41\pm0.02\,.
\end{eqnarray}
Here we only show the major errors stemming from the uncertainties
in the parameter $\eta$ and the cutoff scale $\Lambda$ (see Eq.
(\ref{Lambda})) where we have assigned a 15\% error to $\Lambda_{\rm
QCD}$ and an error of 0.01 to $\eta$. As in Ref.\cite{CCSfsi}, we
have assumed monopole behavior [$n=1$ in Eq. (\ref{FF})] for the
form factor $F(t,m_{D_s})$ and a dipole form ($n=2$) for
$F(t,m_{D_s^*})$. It should be stressed that the estimate of the FSI
contributions is model-dependent as it depends on how we model the
final-state rescattering. In view of this point and the theoretical
discrepancy between PQCD and the topological diagram approach for
the rate of $B\to D_s^{(*)+}D_s^{(*)-}$, it is conceivable that the
actual theoretical uncertainties are considerably larger than those
given in Eq. (\ref{BRtheory}). At any rate, it is evident that the
final-state rescattering contribution to $\bar B_d^0\to J/\psi \phi$
is smaller than the effects of $\omega-\phi$ mixing by two orders of
magnitude. We thus confirm the argument by Gronau and Rosner
\cite{Gronau:2008kk} that a significant enhancement of this mode by
rescattering is unlikely.

\vskip 0.4cm {\bf 4.}~
In this work we have examined the
contributions from photoproduction and final-state rescattering to
$\bar B_d^0\to J/\psi \phi$ and found that the corresponding
 branching ratios are of order $10^{-11}$ and $10^{-9}$, respectively. Hence,
 this decay is dominated by the $\omega-\phi$ mixing effect as advocated by Gronau and Rosner.

\vskip 2.5cm {\bf Acknowledgments}

We are grateful to De-Shan Yang, Cai-Dian L\"{u} and Yang Liu for
useful discussion.
 This research was supported in part by the National
Science Council of R.O.C. under Grant No. NSC97-2112-M-001-004-MY3
for H.Y.C. and by the National Science Foundation under contract
Nos.107-47156 and 10805037 for Y. Li.


\end{document}